\documentstyle{l-aa}
\input{psfig.sty}

\newcommand{\etal}{et al.}
\newcommand{\HI}{\mbox{H\,{\sc i}}}
\newcommand{\HII}{\mbox{H\,{\sc ii}}}

\begin{document}
\thesaurus{03(11.09.1 M\,31; 11.09.4; 09.04.1; 13.09.4)}
\title {Mid--infrared and far--ultraviolet observations of the star--forming 
ring of M\,31\thanks{Based on observations with ISO, an ESA project with 
instruments funded by ESA member states (especially the PI countries: France, 
Germany, the Netherlands and the United Kingdom) and with the participation 
of ISAS and NASA.}}

\author{
        L. Pagani\inst{1}\and
        J. Lequeux\inst{1}\and
	D. Cesarsky\inst{2}\and
        J. Donas\inst{3}\and
	B. Milliard\inst{3}\and
	L. Loinard\inst{4}\and
        M. Sauvage\inst{5}
}
\offprints{james.lequeux@obspm.fr}
\institute{
DEMIRM, Observatoire de Paris, 61 Avenue de l'Observatoire, F-75014
Paris, France
\and
Institut d'Astrophysique Spatiale, Bat. 121, Universit\'e Paris
XI, F-91450 Orsay CEDEX, France
\and
Laboratoire d'Astronomie Spatiale, BP8, Traverse du Siphon, F-13376
Marseille 12 CEDEX, France
\and
IRAM, 300 Rue de la Piscine, Domaine Universitaire, F-38406 St Martin
d'H\`eres CEDEX, France
\and
SAp/DAPNIA/DSM, CEA-Saclay, F-91191 Gif sur Yvette CEDEX, France}
\date{Received 27 July 1998; accepted 23 August 1999}
\maketitle
\begin{abstract}

We present mid--IR images of a 15\arcmin$\times$15\arcmin~ field in
the south--west part of the Andromeda galaxy M\,31 obtained with the
ISOCAM camera (6\arcsec~ pixels) on board ISO. These broad--band
images complement spectro--imaging observations of smaller fields
(Cesarsky et al. \cite{Cesarsky98}).  We also present a 20\arcsec~
resolution far--UV image of a larger field at 200 nm obtained with the
balloon--borne telescope FOCA 1000. These images are inter--compared
and also compared with \HI, CO(1--0) and H$\alpha$ maps.  The mid--IR
emission as seen through wide--band filters centered at 7 and 15
$\mu$m is extremely well correlated with the distribution of neutral
gas as shown by the \HI~ and CO(1--0) maps, while the correlation is
poorer with the distribution of the ionized gas seen through its
H$\alpha$ emission.  There is some correlation with the UV radiation,
but it appears that the contribution of UV photons to the excitation
of the carriers of the mid--IR emission is not dominant in most of
M\,31. The spectro--imaging observations of Cesarsky et
al.  (\cite{Cesarsky98}) show that the mid--IR spectra of several
regions of M\,31, two of which are in the presently studied area, are
dominated by a strong emission band at 11.3 $\mu$m while emission in
the other classical Aromatic Infrared Bands (AIBs) at 6.2, 7.7 and 8.6
$\mu$m is faint or absent. This result is precised, and
we find that the mid-IR spectral variations are not clearly related to
the UV radiation field.  The present observations have important
consequences on our understanding of excitation of the interstellar
mid--IR emission in general. In particular, we conclude that like for M\,31,
excitation in the Galactic cirruses may not be dominated by UV photons
but rather by another mechanism which remains to be identified
(visible photons ?).  The UV
excitation appears to become important when the UV radiation density is of the
order of 2 times that near the Sun.


\end{abstract}
\section{Introduction}
Most of the star formation in the Andromeda galaxy M\,31 is concentrated in 
a ring 10 kpc in radius, but even there the rate of star formation per unit 
area is modest. The far--IR emission of M\,31 is very faint 
for a spiral galaxy 
and its color temperature between 60 and 100 $\mu$m is also particularly low 
(Rice et al. \cite{Rice}; Walterbos \& Schwering \cite{Walterbos87}). The 
South--West 
part of the ring has been studied at many wavelengths, as summarized and 
discussed by Loinard et al. (\cite{Loinard96}, \cite{Loinard98}). This ring 
is in fact the superimposition of several spiral arms which are difficult to 
disentangle from each other owing to the high inclination of the galaxy 
(77\degr). The most prominent of these arms, arm S4 (Baade \cite{Baade}), is 
very similar in its CO line emission to the Carina spiral arm in our Galaxy;
beyond a radius of about 8 kpc (the Solar circle) M\,31 and the Galaxy are 
comparable in their interstellar matter (ISM) content. We will see however 
that the star formation rate per unit surface is smaller than 
in the Milky Way even in the ring. It is considerably smaller in the more 
central parts.

We have imaged various parts of M\,31 with ISOCAM. Four 
3\arcmin$\times$3\arcmin~ fields have been observed with the Circular 
Variable Filters (CVF) in the wavelength range 5.15 to 16.5 $\mu$m
(Cesarsky et al. \cite{Cesarsky98}). The CVF spectra obtained are unlike 
anything that has been observed before from the interstellar medium in our 
Galaxy and in other galaxies. The spectra of the central region and of a 
region in the bulge show a strong, broad emission band at 11.3 $\mu$m while the 
other Aromatic Infrared Bands (AIBs) at 6.2, 7.7 and 8.6 $\mu$m are 
not or only marginally detected. No observation of M\,31 in the 3.3 $\mu$m
AIB exists. Note that a recent ISOCAM filter
imaging of the spiral galaxy NGC\,7331 (Smith \cite{Smith}) appears to show 
a similar situation in the bulge of this galaxy. The emission of the bulge
of NGC\,7331 in 
the LW8 (10.7--12.0 $\mu$m) filter is strong compared to its emission in 
filters around 7 $\mu$m (after the stellar continuum contribution is 
subtracted). In the bulge of M\,31 (Cesarsky et al. \cite{Cesarsky98}), the 
11.3 $\mu$m band has a spatial distribution very different from that of the 
stars 
and is therefore emitted by interstellar rather than circumstellar material. 
The spectrum obtained in a quiet region of the star--forming ring is similar 
to those obtained near the center, while the spectrum of an active 
star--forming region is more comparable to usual Galactic or extragalactic 
AIB spectra, although the 11.3 $\mu$m band is still particularly strong
compared to the other bands. 

In the present paper, we report the results of ISOCAM observations of a
15\arcmin$\times$15\arcmin~ field
in the SW side of the ring made through the LW2 (5.0--8.0 $\mu$m) and
the LW3 (12.0--18.0 $\mu$m) filters. These observations were made in the 
guaranteed time of ISO as parts of a systematic program on interstellar matter 
in the Galaxy, the Magellanic Cloud, M\,31, M\,33 and more distant galaxies, 
in particular a complete sample of galaxies in the Virgo cluster (Boselli et 
al. \cite{Boselli97}, \cite{Boselli98}). The CVF observations mentioned before 
show that these filters were not the best choices for M\,31. The LW8 
(10.7--12.0 $\mu$m) filter which encompasses the 11.3 $\mu$m band would have 
been very useful, but it was too late to 
obtain further observations before the end of the life of ISO. The filter 
observations of the central regions of M\,31 will be described in a 
forthcoming paper.

We also present observations of the same part of M\,31 in the far--UV at
200 nm. These data were obtained with the balloon--borne telescope
FOCA 1000 of the Laboratoire d'Astronomie Spatiale in Marseille, and are 
very complementary to the mid--IR observations.

Sect. 2 describes the ISO observations and their reductions. Sect. 3
is devoted to the observations in the far--UV and their reductions. Sect.
4 compares the morphology of the different emissions observed from far--UV
to radio wavelengths. Sect. 5 discusses the excitation of the mid--IR
emission and Sect. 6 compares the mid--IR emissions in the two
ISOCAM filters LW2 and LW3. Sect. 7 compares the mid-IR emission in
LW2 and LW3 with the CVF observations. Sect. 8 contains the conclusions.

\section{ISO observations and data reduction}

The mid--infrared observations have been made with the 32$\times$32 element
camera (CAM) on board of the ISO satellite, (see Cesarsky et al. 
\cite{CCesarsky} for a complete description). The pixel size was 
6\arcsec$\times$6\arcsec. We obtained for each filter (LW2 and LW3) a square 
9$\times$9--step raster map with a shift (then overlap) of 16 pixels between 
successive positions. The integration time was 2.1 seconds. 50 exposures were 
eliminated at the beginning of each observation in order to insure a better 
stabilization of the detector, which shows remanence of its previous 
illumination history. 20 2.1s exposures were added for each raster 
position. The total useful integration time was about 1 hour per filter.

Data reduction was accomplished using the CAM Interactive Analysis
(CIA) software\footnote{CIA is a joint development by the ESA
Astrophysics Division and the ISOCAM Consortium led by the ISOCAM PI,
C. Cesarsky, Direction des Sciences de la Mati\`ere, C.E.A.,
France}. Most of this analysis is described by Starck et
al. (\cite{Starck}).  The map at each wavelength was dark--current
subtracted using the dark frame model appropriate to the epoch of the
observations. It was then flat--fielded using library flats. Automatic
softwares were used to detect and eliminate the parts of the record
affected by glitches due to the impact of charged particles, and also
to correct for the transient response of the detector when submitted
to changes in incident flux. The new transient correction described by
Coulais \& Abergel (\cite{Coulais}) has been applied, yielding
considerable improvements with respect to the previous method
implemented in the CIA software. In particular, the photometric
accuracy of our maps is now believed to be of the order of 10 \%.
Corrections for field distortion have been applied to images at each
position before combining them in a raster. The final raster images
are of excellent cosmetic quality.  Their resolution is limited by the
6\arcsec~ sampling at short wavelengths and by diffraction at longer
wavelengths, reaching 8\arcsec~ at 15 $\mu$m.  A constant background
corresponding mostly to the zodiacal light, measured in apparently blank
regions of the maps (bottom left and top right corners of Fig. 1) has been subtracted from the LW2 and the LW3
images. The significance of this offset will be discussed later in Sect. 4.

\section{Far--UV observations and reductions}

Far--UV observations of M\,31 have been made in October 1985 during a
balloon flight of the FOCA 1000 ultraviolet telescope of the Laboratoire
d'Astronomie Spatiale. This telescope has a diameter of 40 cm and a focal 
length of 1000 mm. It is equipped with a UV camera consisting of a microchannel 
plate and a Kodak IIaO photographic film. The field of view is 2\fdg3 and 
the angular resolution is 20\arcsec. The bandpass is $\sim$ 15 nm wide 
centered at $\lambda_0 = 200$ nm. This facility is described in Milliard et 
al. (\cite{Milliard}). Six 200 seconds exposure frames centered on the guide 
star HD 3431 have been obtained. Each frame was digitized with a PDS 1010s 
densitometer with a slit size of 25 $\mu$m $\times$ 25 $\mu$m and processed by 
an automatic sequence of programs (Moulinec \cite{Moulinec}) designed to 
convert the photographic density into a linear intensity scale, to correct 
for the detector background and for local differences in sensitivity of the 
microchannel plate. They have then been digitally summed to form the final 
image. The resulting 1640 $\times$ 1640 pixel image is equivalent to a single 
exposure of 1200 s. This image has been corrected for the distorsion of the 
optics to an accuracy $\simeq$ 2\arcsec~ and recentered using stars of the 
Digital Sky Survey to produce Fig. 3 and Fig. 8. The photometric zero point 
has been determined on the basis of 11 stars in common with the previous 
SCAP2000 observation (Donas et al. \cite{Donas}). This calibration gives
fluxes 0.1 mag. fainter than the ANS photometry of several fields of the 
SW part of M\,31 (Isra\"el et al. \cite{Israel}), clearly within the 
uncertainties of both calibrations. The random error on the fluxes is
of the order of 0.35 mag. (Donas et al. \cite{Donas91}).

\section{Morphology of the mid--IR emission and other components of M\,31}

Figure 1 is the LW2 (5.0--8.0 $\mu$m) map superimposed over the Digital Sky
Survey (DSS) image clipped in order to show only stars. 
A number of Galactic stars are visible in the LW2
image and have been used to recenter this image. The LW3 (12.0--18.0 $\mu$m)
image (not shown in full but see Fig. 14, 15 and 16 later) has been itself 
recentered on the LW2 one, having many
features in common. The position accuracy after recentering is of the order
of half a pixel (3\arcsec), similar to that of the UV map.

Figure 2 is similar to Fig. 1 except that the DSS image 
shows fainter levels and that the LW2 image is presented as contours.
It shows the excellent correspondence of the near--IR emission with the 
distribution of large dust grains as revealed by the absorption marks in 
front of the stellar emission of M\,31. The LW3 image being very similar to 
the LW2 one shares this property. A similarity between the LW3 emission and
the distribution of big dust grains has also been seen and discussed by
Block et al. (\cite{Block}) in the case of M\,51. 
Figure 2 also shows that the contribution
of stellar photospheres that dominate the visible image is negligible in the
LW2 filter, at least for the ring of M\,31. In the central parts of M\,31
discussed in a future paper, this is no more true.

\begin{figure}
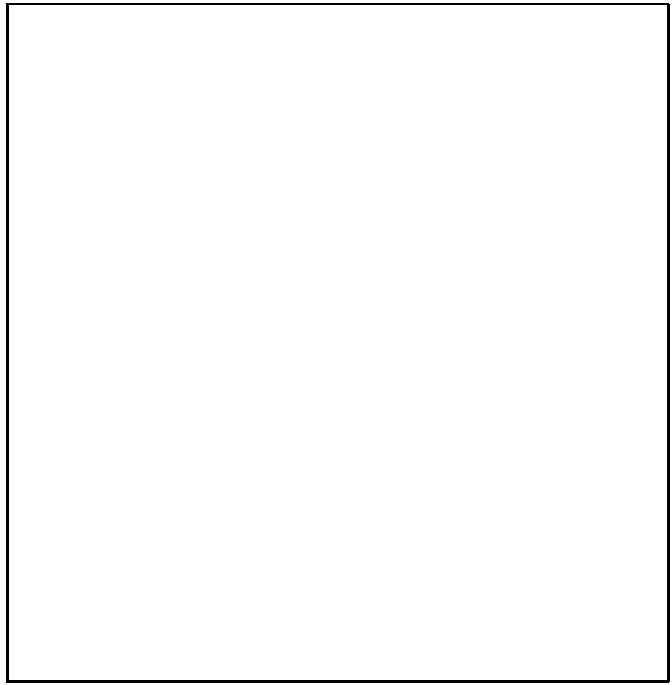

\picplace{9cm}
\label{fig1}
\caption{Map of the SW portion of the ring of M\,31 in the
LW2 filter (5.0--8.0 $\mu$m) (grey scale), superimposed on the
DSS image, clipped such as to show only the stars.
Coordinates are for J2000. The grey scale unit is 1 mJy per 6\arcsec~
pixel, in decimal logarithmic scale.  The two 3\arcmin~ fields
observed with the Circular Variable Filters (CVF) of ISOCAM are
indicated by two squares. Field {\bf c} covers an active star-forming
region while Field {\bf d} corresponds to a quiet region dominated by
molecular gas.  Note the rich OB association NGC\,206 to the West of
Field {\bf d}. This labelling makes reference to Cesarsky et al. (\cite{Cesarsky98}) }
\end{figure}

\begin{figure}
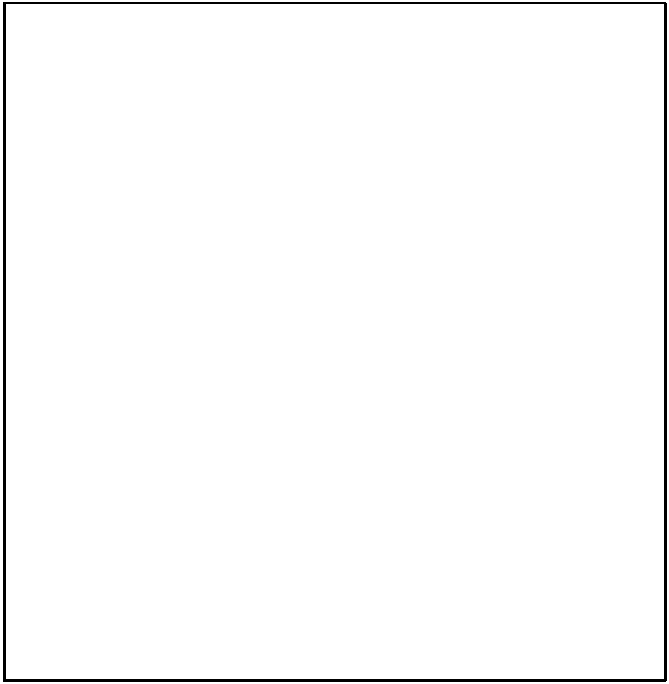

\picplace{9cm}
\label{fig2}
\caption{Map of the SW portion of the ring of M\,31 in the
LW2 filter (5.0--8.0 $\mu$m) (contours), superimposed on the positive,
full--sensitivity DSS image.  Coordinates are for J2000. The first
contour level is 0.4 mJy per 6\arcsec~ pixel then the contours are
from 1 to 6 mJy/pixel by steps of 0.5 mJy/pixel. The levels might be
too low by 0.3 mJy/pixel (see Sect. 4). Note the excellent
correspondence of the near--IR emission with the absorption marks in
front of the stellar emission of M\,31  }
\end{figure}

Figure 3 is the far--UV (200 nm) map (see Sect. 3). 

\begin{figure}
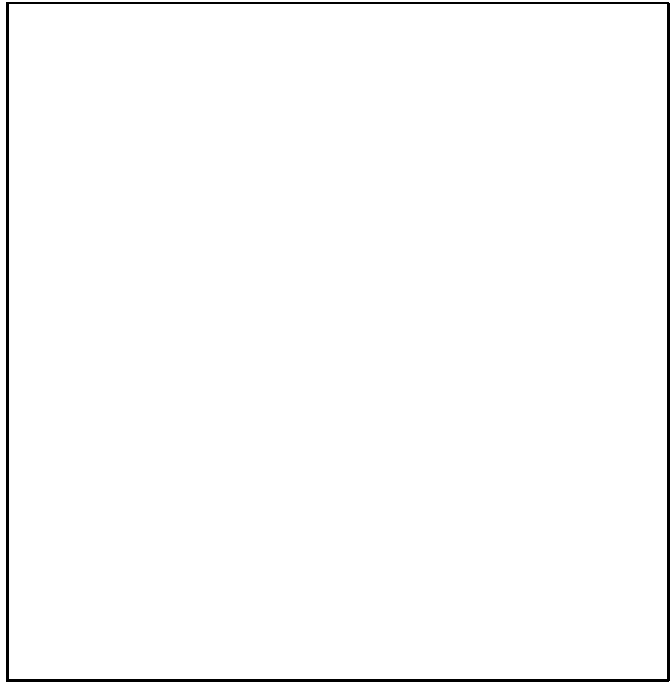

\picplace{9cm}
\label{fig3}
\caption{Map of the SW portion M\,31 in the far--UV at 200 nm obtained
with the FOCA--1000 balloon--borne telescope. The angular resolution is
$20\arcsec$. The color scale unit is 2.1 10$^{-17}$ erg cm$^{-2}$ s$^{-1}$ 
\AA$^{-1}$ arcsec$^{-2}$. The background level as measured outside the galaxy
is 0.116 unit. It has no physical significance. The UV emission is a 
tracer of recent 
($<3~ 10^8$ yr) star formation but radiation also results from scattering 
by dust of the stellar UV emission
}
\end{figure} 

Figures 4 and 5 show the superimpositions of the LW2 
map over the \HI~ and CO maps respectively. The emission seen in the LW2 
filter follows extremely well the distribution of 
the interstellar matter as seen in the \HI~ and the CO lines. 
As we will see later, it is likely that this emission is dominated by the 
AIBs at 6.2, 7.7 and 8.6 $\mu$m. These bands are faint in M\,31 
but the LW2 emission is also quite faint. Tran (\cite{Tran}) has shown that
in general the mid--IR bands are well correlated with \HI, in particular
in photodissociation regions. 
On Fig. 6, we plot the
the intensity in the LW2 filter as a function of the total column 
density of neutral gas N(H) = N(\HI) + 2N(H$_2$), N(H$_2$) being derived 
from the intensity of the CO(1--0) line by:

N(H$_2$) = 1.5 10$^{20}$ I(CO) mol. cm$^{-2}$ (K km/s)$^{-1}$
 
The conversion factor, adapted from Dumke et al. (\cite{Dumke}) and Digel
et al. (\cite{Digel}), is uncertain by at least 50 \%. Our choice is
in agreement with the factor 1.2 10$^{20}$ mol. cm$^{-2}$ (K km/s)$^{-1}$
found by Neininger et al. (\cite{Neininger}). The result of the correlation 
is not very sensitive to its value because most of the 45\arcsec~ pixels used
for building Fig. 6  are dominated by \HI. Note that there is an
error in the \HI~ column densities given by Loinard et
al. (\cite{Loinard96}) which is corrected here and in Loinard et
al. (\cite{Loinard98}). A least--square
fit to the points of Fig. 6 yields:

I(LW2)= $(2.24\pm0.06)$10$^{-22}$ N(H) $-0.23$ mJy pixel$^{-1}$
(H--atom cm$^{-2}$)$^{-1}$ with a correlation coefficient r$^2$ =
0.79. The non-zero intercept is probably not physically
significant. It may be due to the uncertainty on the base level of the
LW2 map and perhaps also of the \HI~ map. From the column density of
\HI~ and H$_2$ at the points chosen to determine the LW2 background
(see Sect. 2), we expect from the correlation slope a flux in LW2 of
0.3 mJy/pixel instead of the assumed value of zero at these
points. After the corresponding correction, this would yield an actual
intercept of 0.07 mJy/pixel which is negligible.  The slope of the
relation is similar within the fairly large errors (at least 30 \%) to
that for the Galactic cirruses as derived from a combination of the
results obtained by Boulanger et al. (\cite{Boulanger96a}), Onaka et
al. (\cite{Onaka}) and Reach
\& Boulanger (\cite{Reach}).  

\begin{figure}
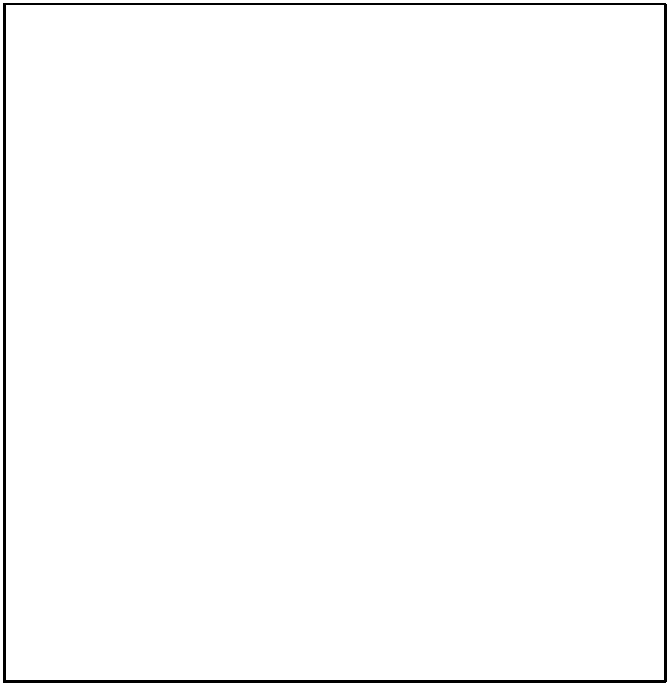

\picplace{9cm}
\label{fig4}
\caption{Map of the SW portion of the ring of M\,31 in the LW2 filter 
(5.0--8.0 $\mu$m) (color). The color scale unit is 1 mJy per 6\arcsec~
pixel, and the scale is truncated at 3 units to obtain a better
display of the faint emission. The contours indicate the 21--cm line
integrated intensity, from data in Brinks \& Shane (1984): contours
with levels from 1000 to 3500 K km/s$^{-1}$ in steps of 500 K
km/s$^{-1}$. The 1000 K km/s$^{-1}$ contour is in red for
clarity. Note that there is an error in the level specification in
Loinard et al. (\cite{Loinard96}). The angular resolution of the
21--cm map is 24\arcsec$\times$36\arcsec.  Note the excellent
correlation between mid--IR and \HI~ emissions, in particular at the
faintest levels. The yellow boxes delineate regions for which the mean
UV intensity has been evaluated, as discussed in the text}
\end{figure}

\begin{figure}
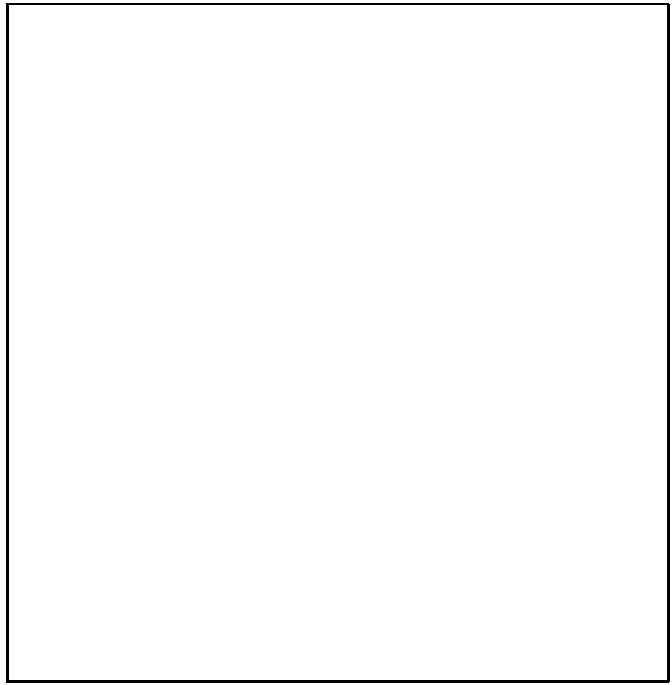

\picplace{9cm}
\label{fig5}
\caption{Map of the SW portion of the ring of M\,31 in the LW2 filter 
(5.0--8.0 $\mu$m) (grey scale), superimposed on the CO(1--0) line map from 
Loinard et al. (1999): contour levels from 2 to 8 K km s$^{-1}$ in steps 
of 1 K km s$^{-1}$.
The angular resolution of the CO map is 45\arcsec. For clarity, the LW2 
image is plotted in a decimal logarithmic scale (unit: 1 mJy per 6\arcsec~ 
pixel).  Note the good correlation 
between the distributions of \HI~ (Fig. 4), CO and mid--IR
}
\end{figure}

\begin{figure}
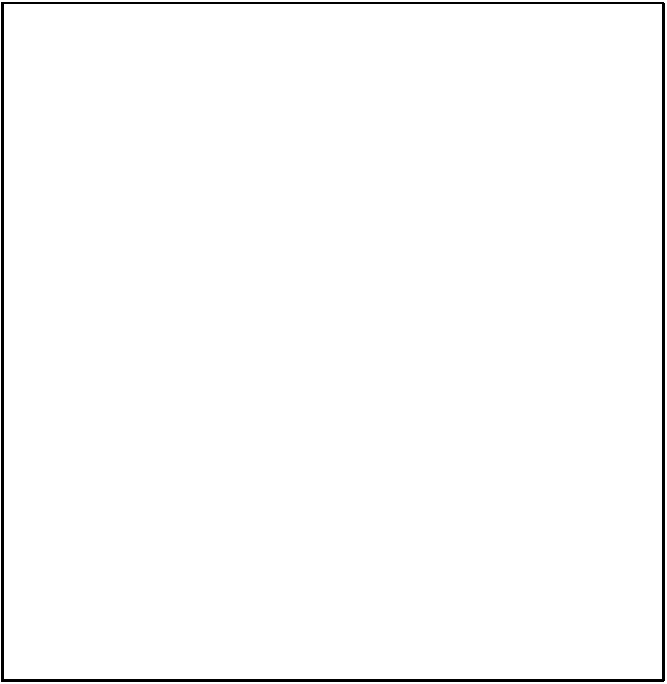

\picplace{9cm}
\label{fig6}
\caption{The intensity in the LW2 filter (5.0--8.0 $\mu$m) as a function
of the total column density of neutral gas N(H) in the SW region of M\,31. 
The size of the pixels used for building this figure is 45\arcsec. The
correlation is excellent. The intercept is discussed in the text
}
\end{figure}

Figures 7 and 8 show superimpositions of the LW2 
map over the H$\alpha$ and far--UV maps respectively. The correlation
between the mid--IR emission and H$\alpha$ is very poor and there is
almost no correlation with the far--UV radiation. There are however
mid--IR peaks associated with \HII~ regions as depicted by H$\alpha$.
In our Galaxy, the emission at 5--8 $\mu$m (the wavelength range of 
the LW2 filter) is strong in photodissociation
regions (PDRs) at the interfaces between \HII~ regions and molecular clouds 
(M\,17: Cesarsky et al. \cite{M17}; Orion bar: Cesarsky et al. \cite{Orion}) 
and at the surfaces of molecular clouds illuminated by UV radiation 
(reflection nebulae like NGC 7023: Cesarsky et al. \cite{NGC7023}). There is 
also emission from the more diffuse interstellar gas (see e.g. Abergel et al.
\cite{rhooph}). However we do not know well the relative importance of these 
various sources at the scale of a galaxy. Our observation of M\,31 gives a 
clear information on this point. The presence of peaks associated
with \HII~ regions probably indicates mid--IR emission 
from interfaces, but there is some ambiguity here because there is also much 
ISM close to the \HII~ regions. In any case, the mid--IR emission of the 
interfaces cannot be a large fraction of the total, since there is much 
diffuse mid--IR emission in M\,31 far from H$\alpha$ emission regions. We 
will see later (fig. 11) that only very few localized regions in M\,31 have
a LW3/LW2 intensity ratio (in Jy) larger than 1, a characteristic of strong
PDRs.   

Particularly interesting is the vicinity of the main OB association of 
M\,31, NGC\,206 = OB\,78 (see Fig. 1). 
The stellar 
winds and perhaps supernova explosions in this association have eroded a hole 
in the interstellar medium 
while triggering some secondary star formation in the resulting
shell (Fig. 1 and 7). The UV emission of the association itself is very 
strong (Fig. 3 and 8, see also Hill et al. \cite{Hill}). The Lyman continuum 
photons from the OB stars of this association (Massey et al. \cite{Massey} 
and references herein) ionize the inner shell as it can be seen clearly in 
H$\alpha$ (Fig. 7). Still the mid--IR flux is not higher in this region than 
in inactive regions with similar column densities of gas. {\it This confirms 
that the main factor which determines the mid--IR flux is the amount of gas, 
except in the immediate vicinity of regions where massive stars have 
just formed}. Several such regions are well visible in the shell around 
NGC 206 (Fig. 7), each one with its 
own mid--IR peak.

We conclude from these morphological considerations that in the ring of 
M\,31 the emission in the LW2 filter (5.0--8.0 $\mu$m) is dominated by the 
diffuse ISM bathed in the general interstellar field. This confirms the 
result of the analysis of IRAS data from the solar neighbourhood by Boulanger 
\& P\'erault (\cite{Boulanger88}). They showed that most of the interstellar 
infrared emission comes from ISM not associated with current star 
formation. Their result actually applies to the far--IR emission but since 
there is a good correlation between the emission in the 12 $\mu$m IRAS filter 
and that in the filters at longer wavelengths
it is likely to be true for the mid--IR emission as well.

\begin{figure}
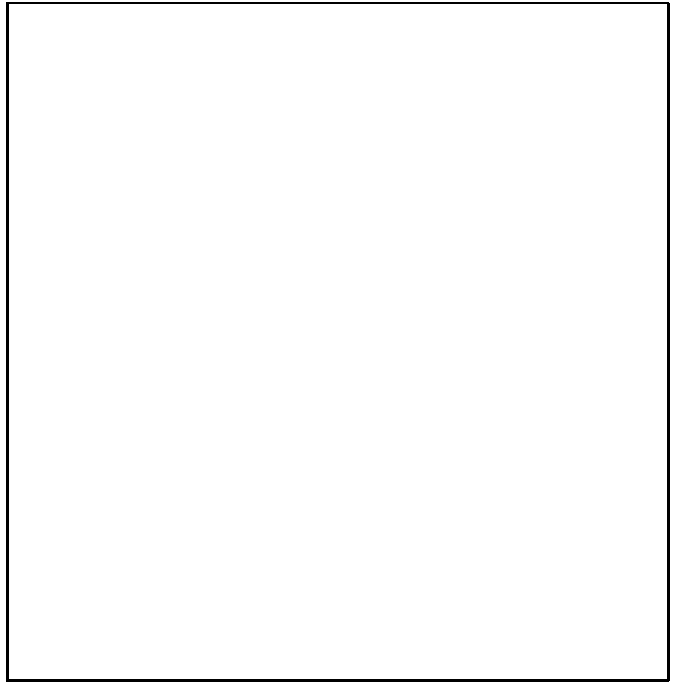

\picplace{9cm}
\label{fig7}
\caption{Map of the SW portion of the ring of M\,31 in the LW2 filter 
(5.0--8.0 $\mu$m) (clipped grey scale), superimposed on an image (contours)
kindly communicated by N. Devereux (see Devereux et al. 1994). The
H$\alpha$ contours are approximately 5 10$^{-16}$, 10$^{-15}$ and 3
10$^{-15}$ erg cm$^{-2}$ s$^{-1}$ per 2\arcsec~ pixel. The unit for the
LW2 image is 1 mJy per 6\arcsec~ pixel. The H$\alpha$ emission is
limited to the outer side of the ring of interstellar matter as
defined by the \HI, CO and mid--IR emission. H$\alpha$ emission
surrounds the association NGC\,206}
\end{figure}

\begin{figure}
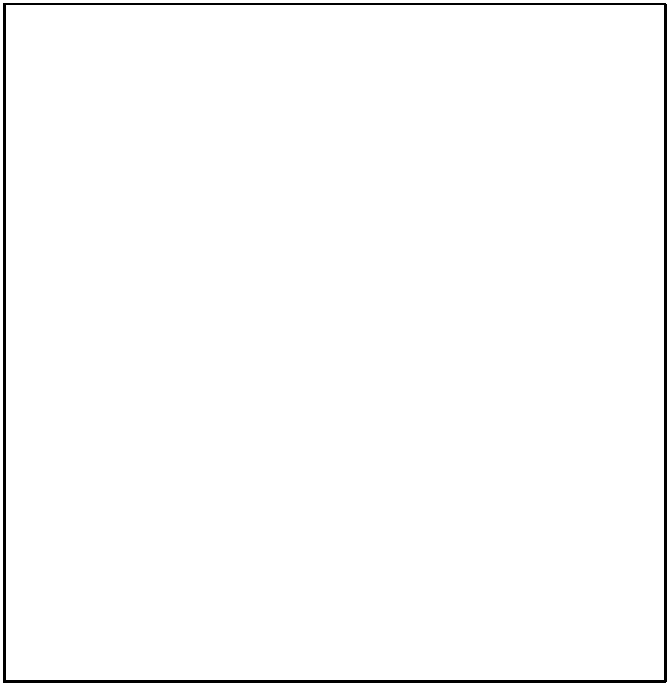

\picplace{9cm}
\label{fig8}
\caption{Map of the SW portion of the ring of M\,31 in the the far--UV at 
200 nm (contours) superimposed on the LW2 filter (5.0--8.0 $\mu$m) map (color).
The UV contours are at 0.15, 0.2, 0.3, 0.4, 0.5, 1.0 and 2.0 in units of 
2.1 10$^{-17}$ erg cm$^{-2}$ s$^{-1}$ \AA$^{-1}$ arcsec$^{-2}$. The 
background is at 0.116 unit. The LW2 filter color scale unit is 1 mJy/pixel.
It is truncated at 3 units for clarity. 
A few OB associations give a very strong UV emission, in particular NGC 206.  
Most of the diffuse UV emission comes from outside the ring of 
interstellar matter as defined by the \HI, CO and mid--IR emission. The UV 
within the ring is affected by extinction. The extended UV emission is 
probably in part light scattered by dust in the diffuse ISM
}
\end{figure}

\section{The excitation of the mid--IR emission in M\,31}

The mid--IR band emission is generally considered to be excited
by individual UV photons that transiently heat these particles to very high 
temperatures. However Uchida et al. (\cite{Uchida}) found a ``normal''
Galactic AIB spectrum in the reflection nebula vdB\,133, which is illuminated
by a relatively cool binary star, and noticed that this poses a problem for 
the emission mechanism of the mid-IR bands. This is amply confirmed by our 
observations of M\,31, as we will demonstrate now. 

\begin{figure}
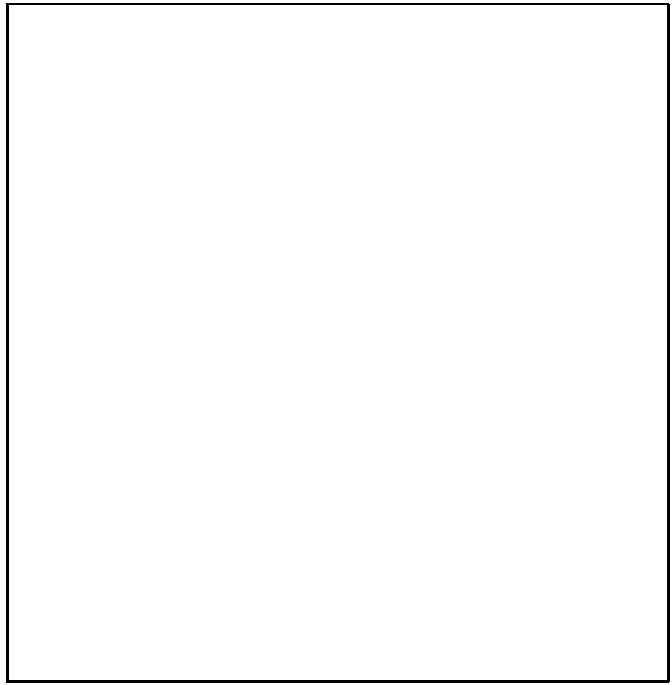

\picplace{9cm}
\label{fig9}
\caption{The intensity in the LW2 filter (5.0--8.0 $\mu$m) normalized by
the total column density of neutral gas in the SW region of M\,31, as
a function of the uncorrected intensity of the far--UV light at 200 nm. 
The size of the pixels used for building this figure is 45\arcsec. There
is only a weak correlation between the LW2 flux per H atom and the intensity
of the UV field which appears above a threshold value of $\simeq$ 2
10$^{-18}$ erg cm$^{-2}$ s$^{-1}$ \AA$^{-1}$ arcsec$^{-2}$ 
}
\end{figure}

The correlation between the mid--IR and the far--UV emission in M\,31 (Fig. 8)
is apparently extremely poor.  The lack of correlation must be due
to some extent to the strong extinction in the gas and dust ring, given
the large inclination of M\,31 on the line of sight. This extinction
affects very much the UV radiation, but little the mid--IR.
 
To get rid of the problem of
extinction, we compare the ratios along the same 21--cm line brightness 
isophote of 1000 K km s$^{-1}$ on each side of the 10 kpc ring
(see Fig. 4, yellow boxes 1 and 2). The column density of H atoms in H$_2$ is
estimated to be 4.5 10$^{20}$ cm$^{-2}$ in box 1 and 3.8 10$^{20}$
cm$^{-2}$ in box 2 compared to 1.8 10$^{21}$ cm$^{-2}$ for \HI. In the 
Appendix we give estimates of the far--UV/visible interstellar intensity 
ratio per unit wavelength in various parts of M\,31. We find
I$_{\lambda}$(200 nm)/I$_{\lambda}$(550 nm) $\simeq 0.20$ on the outer 
side (box 2) which is brighter in UV (see Fig. 8) and 0.05 on the
inner side (box 1). This difference cannot be due to extinction since
the column densities of interstellar matter are the same for the two boxes. 
The intensity in the LW2 filter is almost the same for the two regions
(respectively 0.23 and 0.28 mJy/pixel plus a possible offset of about 0.3
mJy/pixel). 
This example shows beyond
any doubt that the mid--IR intensity is almost independent of the UV radiation 
field in the observed region of M\,31. This 
suggests that the AIBs in M\,31 are in general not primarily excited by the 
far--UV photons.

The far--UV/visible interstellar intensity ratios on both sides of the
ring of M\,31 are affected by the same amount of extinction and a model
is required if one wishes to obtain dereddened ratios. This is done in the 
Appendix where we show that the dereddened ratios are roughly 1.55 times 
larger than the observed ones. 
For comparison, I$_{\lambda}$(200 nm)/I$_{\lambda}$(550 nm) 
$\simeq 0.65$ in the Solar neighbourhood. Thus on both sides of the ring
the I$_{\lambda}$(200 nm)/I$_{\lambda}$(550 nm) ratio is smaller than near 
the Sun, and it is clear that in this case the UV radiation has a negligible
effect to excite the mid--IR emission. In the Appendix, we also show
that the UV/visible ratio in the reflection nebula vdB 133 observed by 
Uchida et al. (\cite{Uchida}) is of the order of 0.4, slightly
smaller than in the Solar neighbourhood but definitively larger than
in quiescent regions of the ring of M\,31. Thus our observations of M\,31 
yield an even more convincing case for the fact that excitation of the 
bearers of the AIBs (PAHs?) does not require UV radiation. Excitation
might be by visible photons. In this case the emitters, if they are PAHs, 
must be smaller than usually assumed since single--photon
excitation by photons some 5 times less energetic than considered usually
should bear them transiently to a temperature of several hundred degrees.
Alternatively, the AIB excitation may be linked in some way to \HI~ given 
the excellent correlation between mid--IR and \HI. A possibility could be that
formation of H$_2$ molecules by combination of H atoms on the AIB carriers 
supplies their heating (Papoular \cite{Papoular}). However it is easy to 
see that this mechanism fails by two orders of magnitude in the case of 
Galactic PDRs and cirruses (F. Boulanger, private communication). This
is probably also the case for M\,31.

There is however an underlying correlation between the mid-IR flux {\it per
H atom} and the UV flux. This is illustrated by Fig. 9, where we have 
plotted the intensity in the LW2 band normalized by the total column 
density of the neutral gas N(H) as a function of the intensity
at 200 nm. The intensity in the LW2 band has been corrected from the non--zero
intercept seen on Fig. 6 by adding to it 0.3 mJy/pixel before dividing by 
N(H) (see Sect. 4). From Fig. 9 we see
that UV excitation becomes important only above an uncorrected surface
brightness I$_{UV}$ $\simeq$ 2 10$^{-18}$ erg cm$^{-2}$ s$^{-1}$
\AA$^{-1}$ arcsec$^{-2}$. The least--square regression line has a slope of 10$^{-22}$ mJy 
pixel$^{-1}$ cm$^{-2}$ per 10$^{-18}$ erg cm$^{-2}$ s$^{-1}$
\AA$^{-1}$ arcsec$^{-2}$ for I$_{UV} >$ 2 10$^{-18}$ erg cm$^{-2}$ s$^{-1}$
\AA$^{-1}$ arcsec$^{-2}$, mostly determined by a few UV--bright
regions. This would determine the amount of excitation due to the UV
radiation if one could correct the UV flux for extinction, a very
difficult task for UV-bright regions. In any case, the extinction
correction is likely to decrease strongly the slope of the regression
line.

The weakness of the mid-IR/UV correlation has important
consequences. It seems to imply that at low UV intensities UV
excitation of the LW2 emitters is unimportant with respect to another
excitation mechanism that remains to be determined. The dereddened UV
surface brightness corresponding to the threshold for UV excitation is
of the order of 10$^{-17}$ erg cm$^{-2}$ s$^{-1}$
\AA$^{-1}$ arcsec$^{-2}$ (see Appendix). It is
easy from data in the Appendix to calculate that the true (unreddened) UV
surface brightness of the Galactic disk near the Sun, assumed to be
seen from the same inclination of 77\degr~ as M\,31, would be $\simeq$
5 10$^{-18}$ erg cm$^{-2}$ s$^{-1}$ \AA$^{-1}$ arcsec$^{-2}$ at 200
nm. This is less than the above threshold and suggests that the mid-IR
emission of the Galactic cirruses is also only marginally excited by
UV photons. This may explain why the ratio
between the LW2 intensity and N(H) is so similar in M\,31 and for Galactic
cirruses in spite of the difference in their UV fluxes. A UV radiation
density 2 times larger than that near the Sun appears to be needed for
UV excitation to become appreciable in the disks of M\,31 and of the Galaxy.

\section{The LW3(12.0--18.0$\mu$m)/LW2(5.0--8.0$\mu$m) band ratio}

In the previous sections, we have only discussed the LW2 observations. The LW3
(12.0--18.0 $\mu$m) image (not shown) is at first sight very similar to
the LW2 (5.0--8.0 $\mu$m) one. We show on Fig. 10 the pixel--to--pixel 
correlation between the brightnesses in the two maps. A Hanning smoothing 
has been applied to both the LW2 and the LW3 images in order to obtain similar 
angular resolutions at both wavelengths. The correlation is excellent. The 
two-dimension regression line gives 

I(LW3) = 0.82 I(LW2) - 0.10 

with a correlation coefficient r$^2$ = 0.72.  The units are mJy/pixel
for both intensities.  The overall similarity between the LW3, the LW2
map and the neutral gas indicates that the excitation mechanisms in
the two wavelengths ranges must be related to each other and to the
neutral ISM.

\begin{figure}
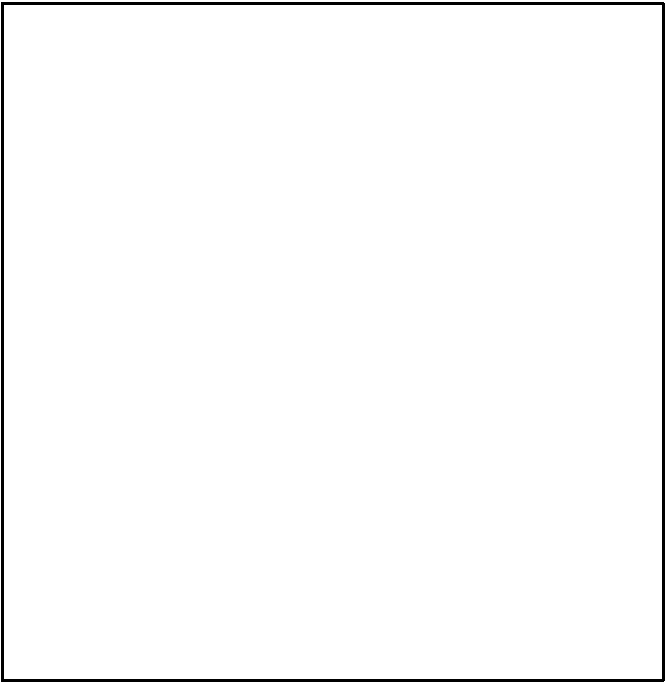

\picplace{9cm}
\label{fig10}
\caption{The intensities measured in the LW3 (12.0-18.0 $\mu$m) as a function
of the intensities in the LW2 (5.0--8.0 $\mu$m) filter. A Hanning smoothing 
has been 
applied to the LW2 and LW3 maps in order to obtain similar angular resolutions 
at both wavelengths. 
Notice the bifurcation of the correlation at large 
intensities. The deviating points forming an almost horizontal sequence below 
the main correlation correspond to stars
}
\end{figure}

The I(LW3)/I(LW2) ratio is surprisingly similar to that in more ``normal''
galaxies. For comparison, in the region of the $\rho$ Ophiuchi cloud mapped by 
Abergel et al. (\cite{rhooph}) this ratio is between 0.6 and 1.0. It is
somewhat larger than 0.8 everywhere in M\,51 (Sauvage et al. 
\cite{M51}) or in NGC\,6946 (Helou et al. \cite{NGC6946}). Integrated over 
whole normal galaxies, the I(LW3)/I(LW2) ratio is always larger than 0.5 
except for a few irregular galaxies (Boselli et al. \cite{Boselli97}, 
\cite{Boselli98}).

\begin{figure}
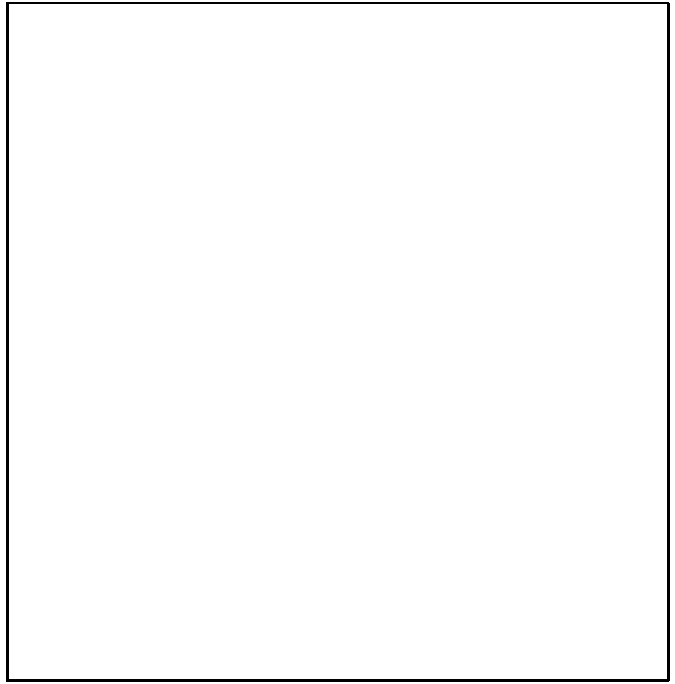

\picplace{9cm}
\label{fig11}
\caption{A partial map of the ratio between the intensities measured in the 
LW3 (12.0--18.0 $\mu$m) and in the LW2 (5.0--8.0 $\mu$m) filters. 
A Hanning smoothing 
has been applied to the LW2 and LW3 maps in order to improve the 
signal--to--noise ratio and to have approximately the same angular resolution 
at both wavelengths. Only the points for which the intensities are 
larger than 4 times the r.m.s. noise have been used. The H$\alpha$ map is
superimposed (contours from 5 10$^{-16}$ to 5 10$^{-15}$ by steps of 5
10$^{-16}$ erg cm$^{-2}$ s$^{-1}$ per 2\arcsec~ pixel).  Only a part of the total field is depicted, 
showing a few ``hot spots'' with a large
LW3/LW2 intensity ratio. There are no such conspicuous hot spots in the 
rest of the surveyed region
}
\end{figure}

In spite of the overall similarity between the LW2 and LW3 maps, there are 
interesting differences. The existence of localized regions where the LW3/LW2 
intensity ratio deviates from the mean at relatively large intensities is 
visible on Fig. 10. The points with a low LW3 intensity with respect to
the LW2 one correspond to stars. For these points one has approximately
I(LW3)/I(LW2) $\simeq$ 0.25, as expected for a Rayleigh--Jeans blackbody law.
There are also a few points for which I(LW3)/I(LW2) is large. 
Fig. 11 is an enlarged map of the intensity
ratio in a part of the M\,31 ring, on which a few such ``hot spots'' can 
be seen. A possible explanation is that if the radiation field becomes 
large enough, then very small 3-dimensional grains 
contribute to the emission in the LW3 filter. This has been discussed e.g. by 
Cesarsky et al. (\cite{M17}). Curiously,
the hot spots are close to \HII~ regions, but not coinciding with them.
It is likely that we see the interfaces between the \HII~ regions and
molecular clouds, or material heated by young, massive stars embedded
in molecular clouds. The angular resolution of our maps, in particular
the CO one, does not allow to check this point. Near--IR observations will
be very useful to ascertain the nature of the hot spots.

\section{The nature of the emission in the LW2 and LW3 filters: comparison
with CVF observations} 

CVF observations are available for Fields c and d of Fig. 1. Their global
spectrum is shown in Fig. 3c and 3d of Cesarsky et al. (\cite{Cesarsky98})
respectively.
Field d is a quiescent region with little star formation and its spectrum
is dominated by a broad band centered at 11.3 $\mu$m, with only marginal
emission in the 6.2, 7.7 and 8.8 $\mu$m bands. We have built a map of the
11.3 $\mu$m feature in this field. It shows some relation with
the LW2 and LW3 maps, but it is very noisy due to the faintness of the 
emission and no clear conclusion can be extracted from it. Field c is 
a more active region and its mid--IR emission is stronger. Its global CVF 
spectrum shows all the classical AIBs although the 11.3 $\mu$m is 
relatively very strong. We will now discuss its mid--IR emission in detail.

We performed a further reduction of the CVF observation of Field c. It 
has been reprocessed with the new transient correction algorithm (Coulais
\& Abergel \cite{Coulais}), and recentered using a star which is visible
on the LW2 map and in the short--wavelength channels of the CVF. We then
defined a background emission spectrum using pixels considered as free from 
emission in the AIBs. This spectrum, which is essentially the zodiacal 
light spectrum, has been subtracted from the spectra of the individual pixels
in order to produce differential spectra showing more clearly the AIBs.
Note that the continuum is partly lost in this process because the response
of the detector to the zodiacal light is not uniform due to reflections
between the CVF and the detector, and not yet well calibrated. 
We have then produced maps in the 6.2 $\mu$m
band and in the 11.3 $\mu$m one, as well as average differential spectra
in the regions emitting sufficiently strongly in these bands.

Figure 12 compares the average differential CVF spectra obtained with three 
different selections of pixels: those with a sufficiently strong emission
at 6.2 $\mu$m but a weak emission at 11.3 $\mu$m, those pixels in the opposite
case and those pixels with a relatively strong emission at both 6.2 and
11.3 $\mu$m. This illustrates the diversity of spectra found in the field,
which go from ``normal'' AIB spectra similar to those in our Galaxy
to spectra with a relatively strong, broad 11.3 $\mu$m
feature and very little emission in the 6.2, 7.7 and 8.6 $\mu$m bands. 

Figure 13 is the CVF map in the 6.2 $\mu$m feature, superimposed over the
LW2 map. We see on this figure that the two distributions are similar, with
a relatively strong peak in the direction of the \HII~ region PAV78\,159 
(Pellet et al. \cite{Pellet}) = BA\,1-313 (Baade \& Arp \cite{BA}). This shows
that the emission in the LW2 filter is due to the 6.2 $\mu$m and 
probably also the 7.7 $\mu$m AIBs, and possibly to an associated continuum.
It is unfortunately not possible at the present stage to simulate the flux 
in the LW2 filter from the CVF data, due to the partial loss of the continuum
level in the CVF reductions.

Figure 14 contains the CVF map in the 11.3 $\mu$m feature, superimposed over 
the LW3 map. The LW3 map is qualitatively similar to the LW2 map displayed on
Fig. 13, the most conspicuous difference being the expected absence of the
Galactic star in the LW3 map. These differences can be seen on the partial map
of Fig. 16. Both filter maps show only a loose similarity 
to the 11.3 $\mu$m band map. We also display on the right of the figure
differential average spectra for the pixels showing emission in the 11.3
$\mu$m band, in three strips of the CVF field. This shows that the 11.3
$\mu$m emission regions of the lower--left band, which are not conspicuous 
in the filter maps, are real. Clearly
the LW3 emission has little to do with the 11.3 $\mu$m feature (note that
this feature is {\it not} included in the passband of the LW3 filter). This 
emission is essentially a continuum. Usually
one attributes this continuum to the emission of very small grains heated
in semi--equilibrium by the absorption of (mainly UV?) photons (D\'esert
et al. \cite{Desert}; Cesarsky et al. \cite{M17}). However this interpretation
is obviously disputable here.

It is interesting to precise in this respect that the contribution of the 
11.3 $\mu$m feature to the emission in the IRAS 12 $\mu$m filter 
(8.0--15.0 $\mu$m) is minor. The average IRAS brightness at 12 $\mu$m in 
Field c is $\simeq$ 0.7 mJy per 6\arcsec~ pixel from Walterbos \& Schwering 
(\cite{Walterbos87}) or Xu \& Helou (\cite{Xu}), of which at most 20 \%
comes from the 11.3 $\mu$m band. The rest must be essentially continuum 
emission, to be compared with an average emission of $\simeq$ 1 mJy/pixel 
in the LW3 filter (12.0--18.0 $\mu$m). Although the result of this 
comparison is encouraging, it would be premature to try to extract the 
spectral slope of the continuum from these very rough numbers.  

\begin{figure}
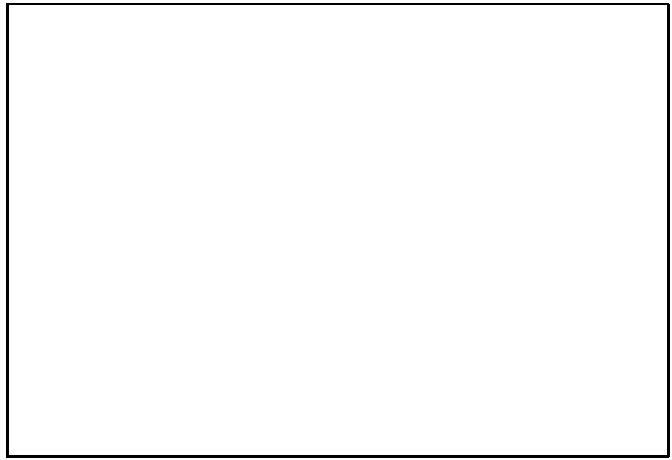

\picplace{6cm}
\label{fig12}
\caption{Average differential spectra in Field c of Fig. 1, an active 
star--forming region of the ring of M\,31, with three 
different selections of pixels. The middle and top spectra are
shifted vertically for clarity. Top: pixels strong in the 11.3 $\mu$m band 
and weak in the 6.2 and 7.7 $\mu$m ones. Middle: pixels strong in all three 
bands. Bottom: pixels strong in the 6.2 $\mu$m band and weak in the 11.3
$\mu$m one. The latter spectrum resembles the usual Galactic AIB spectra. 
A continuum obtained from the pixels with no visible emission in any of the
bands has been subtracted. Note that the continuum level and shape are 
lost in the process, as explained in the text
}
\end{figure}

\begin{figure}
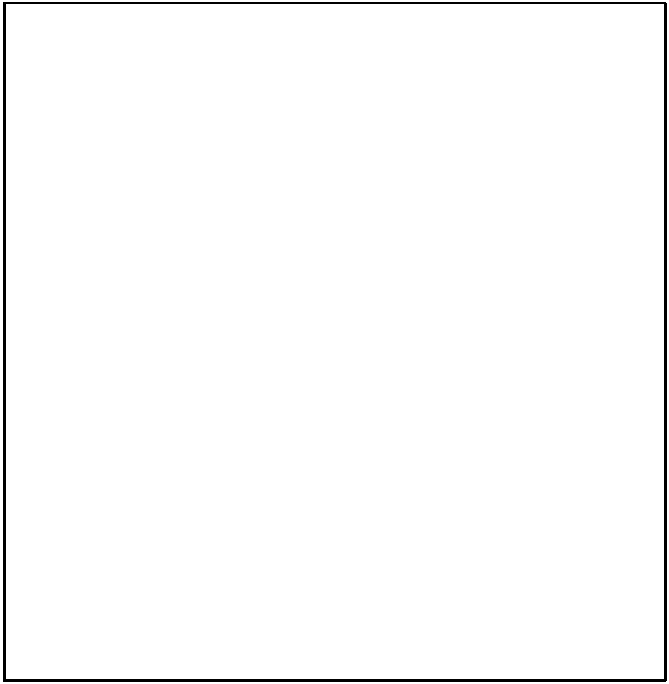

\picplace{9cm}
\label{fig13}
\caption{Map of Field c of Fig. 1, an active star--forming region of the 
ring of M\,31 in the integrated 6.2 $\mu$m emission band (contours)
superimposed on the LW2 filter (5.0--8.0 $\mu$m) image (grey scale, in units 
of 1 mJy per 6\arcsec~ pixel). Coordinates are J2000. The LW2 point source 
at 00h 41m 05s, 40\degr~ 36\arcmin~ 40\arcsec~ is a red Galactic star which
has been used to recenter the CVF 6.2 $\mu$m image on the LW2 one.
The field of the CVF is indicated by the dashed lines. Note the good 
correspondence between the two images, suggesting that the LW2 image is 
dominated by the AIBs
}
\end{figure}

\begin{figure}
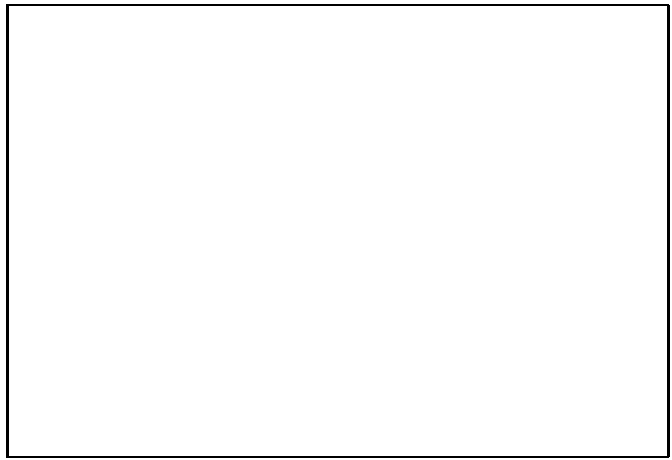

\picplace{6cm}
\label{fig14}
\caption{Map of Field c of Fig. 1, an active star--forming region of the 
ring of M\,31, in the integrated 11.3 $\mu$m emission band (contours)
superimposed on the LW3 filter (12.0--18.0 $\mu$m) image (grey
scale). The average differential spectra for the 11.3 $\mu$m--strong
pixels of three strips in the CVF image are displayed on the right of
the figure. Note in particular that the 11.3 $\mu$m emission regions
of the lower--left strip, which do not correspond to strong peaks in
the LW3 map, do not produce a detectable emission in the 6.2, 7.7 and
8.6 $\mu$m bands }
\end{figure}

\begin{figure}
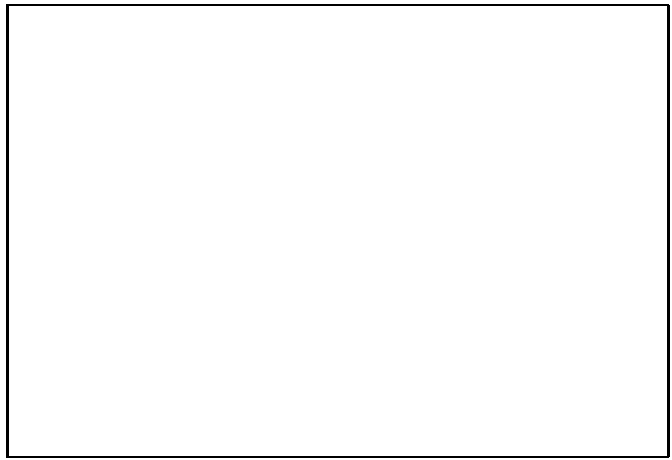

\picplace{6cm}
\label{fig15}
\caption{Map of Field c of Fig. 1, an active star--forming region of the 
ring of M\,31 in the LW3 filter (12.0--18.0 $\mu$m) (color; unit = 1 
mJy/pixel). The white contours are for the 21--cm line integrated 
intensity, from data in Brinks \& Shane (1984): levels from 
1000 to 3500 K km/s$^{-1}$ in steps of 500 K km/s$^{-1}$. 
The red contours indicate the intensity of the CO(1--0) line map from 
Loinard et al. (1999): contour levels 2 to 9 K km s$^{-1}$ in steps of 
1 K km s$^{-1}$. The angular resolution of the 21--cm map is 
24\arcsec$\times$36\arcsec, and that of the CO map is 45\arcsec. The 
correspondence between the LW3 map and the distribution of gas is good as
far as one can judge given the very different angular resolutions,
except for the main LW3 peak which coincides with a \HII~ region
}
\end{figure}

\begin{figure}
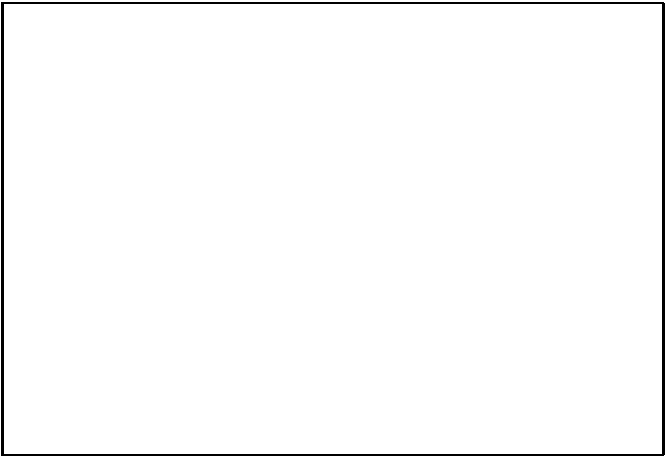

\picplace{6cm}
\label{fig16}
\caption{Map of Field c of Fig. 1, an active star--forming region of the 
ring of M\,31 in the LW3 filter (12.0--18.0 $\mu$m) (color; unit = 1
mJy/pixel). The black contours are for the LW2 image (levels 1 to 4
mJy/pixel by steps of 1 mJy/pixel). The white contours are H$\alpha$
isophotes (levels 0.5, 1.5, 2.5, 5 and 7.5 10$^{-15}$ erg cm$^{-2}$
s$^{-1}$ per 2\arcsec~ pixel). The red contours are UV isophotes at 200
nm (levels 0.15, 0.175, 0.2 to 0.6 by 0.1, in units of 2.1 10$^{-17}$
erg cm$^{-2}$ s$^{-1}$ \AA$^{-1}$ arcsec$^{-2}$; the background is at
0.116 unit).  The main \HII~ regions (H$\alpha$ peaks) are in order of
increasing right ascension PAV78\,147, 152, 159, 157 (south of 159)
and 161 (Pellet et al. 1978). Coordinates are J2000.  The main UV peak
at 00h 40m 55s, 40\degr~ 36\arcmin~ 10\arcsec~ is a hot star or
stellar cluster, probably belonging to M\,31. It coincides with the 12
mag. star BGH88\,40 2579 (Berkhuijsen et al. 1988) but is more
probably to be identified with the UV object HIB95\,82-12 (Hill et al.
1995) in spite of a position difference of 14\arcsec. Another peak at
00h 40m 50s, 40\degr~ 37\arcmin~ 20\arcsec~ is very close to the
emission--line object MLA93 \,270 (Meyssonnier et al. 1993) and to the
red 13th mag. star BGH88\,40\,2486 (Berkhuijsen et al. 1988) so that
the identification is unclear.  Note the excellent correspondence
between the LW2 and LW3 images at faint levels, but also the
differences at higher intensities.  There is little mid--IR emission
in the UV--rich part of the field, and vice--versa. There are however
LW2 and LW3 peaks in the direction of the
\HII~ regions, the strongest peak coinciding with PAV78\,159 (Pellet et al. 
1978)
}
\end{figure}

In order to clarify the situation in Field c, we present on Fig. 15 and
16 comparisons of the ISO filter maps with maps of other quantities.
On Fig. 15, the HI column density and the CO(1--0) line intensity contours
are superimposed on the LW3 image. We note the good overall correspondence 
between the mid--IR and the distribution of gas that we have already seen 
in the full maps of Fig. 4 and 5, especially at faint levels. However the 
strongest LW3 feature does not correspond to a peak in the gas distribution, 
but rather to the \HII~ region PAV78\,159. Fig. 16 compares the LW3 image with the 
LW2, H$\alpha$ and UV images. It shows that there are quantitative 
differences between the LW3 and the LW2 images. The correspondence
is excellent at the faint levels but not so good at higher levels. The
LW2 peak at 00h 40m 50s, 40\degr~ 37\arcmin~ 20\arcsec, not visible
on the LW3 image and which coincides with a UV source, is a star. The
\HII~ regions seen in H$\alpha$ generally coincide with peaks in both the 
LW2 and LW3 maps as expected, while the correspondence is almost inexistent
with the UV. We already remarked these properties at the scale of the
ring of M\,31. Finally
we note by comparing Fig. 14 and Fig. 16 that the regions where the 11.3 
$\mu$m AIB dominates over the other AIBs do not seem poorer in UV.  

\section{Conclusions}

In this paper, we have compared maps of the SW part of M\,31 at different 
wavelengths. In particular the mid--IR maps obtained with ISOCAM with the
LW3 (12.0-18.0 $\mu$m) and the LW2 (5.0--8.0 $\mu$m) filters at a resolution
of 6\arcsec$\times$6\arcsec~ have been compared with lower--resolution
\HI~ and CO maps, with a higher--resolution H$\alpha$ map and with a far--UV
map at 200 nm with a resolution of 20\arcsec~ presented here for the first
time. We also built smaller maps in the 6.2 $\mu$m and in the 11.3 $\mu$m AIB 
from ISOCAM CVF observations
in a relatively active field of the star--forming ring, and compared them 
with other observations. The results of this study can be summarized as 
follows.\\

\begin{itemize}
\item The mid--IR emission is very well correlated with the column density 
of gas (atomic + molecular), but little with the H$\alpha$. 
We conclude from this observation that the emission is globally dominated 
in M\,31 by the diffuse ISM rather than by ISM at the interfaces between 
\HII~ regions and neutral clouds.\\

\item The mid--IR emission shows very little correlation with the far--UV
radiation. This cannot entirely be due to extinction of the far--UV in
regions emitting in the mid--IR. The existence of regions with similar
gas column densities hence similar extinctions and with identical
mid--IR emissions but very different UV intensities shows that the AIB
carriers in M\,31 are not primarily excited by UV radiation except
near star--formation centers. This confirms and extends the
observation by Uchida et al. (\cite{Uchida}) of a Galactic reflection
nebula with little UV showing a ``normal'' AIB spectrum. In M\,31 like
in this case, the excitation of the AIBs is dominated by something
else than UV radiation, probably by photons in the visible.  There is
however some contribution of the UV to the excitation of the mid--IR
emission that becomes apparent when the emission per H atom is plotted
as a function of the UV flux (Fig. 9). In the disks of M\,31 and of
the Galaxy the UV excitation becomes
important when the UV radiation density is of the order of 2 times
that near the Sun. This explains why the 5.0--8.0 $\mu$m emission per
H atom does not differ by more than 30 \% in M\,31 and Galactic
cirruses in spite of the differences in the UV radiation density.\\

\item The emission in the LW2 band (5.0--8.0 $\mu$m) is probably dominated
by the AIBs at 6.2 and 7.7 $\mu$m, which whatever the emission mechanism
are believed to be heated transiently by some quantum mechanism. The
emission in the LW3 band (12.0--18.0 $\mu$m) is dominated by a continuum. 
In our Galaxy, this continuum has often be attributed to
emission by very small carbonaceous grains heated in semi-equilibrium
by UV photons. The ratio between the intensities in these two bands
I$_{\lambda}$(12.0--18.0 $\mu$m)/I$_{\lambda}$(5.0--8.0 $\mu$m)
is very uniform over the observed part of M\,31, of the order of 0.8,
except near star--formation regions. This points to very related excitation
mechanisms.\\

\item There are however a few localized regions with a larger
I$_{\lambda}$(12.0--18.0 $\mu$m)/I$_{\lambda}$(5.0--8.0 $\mu$m) ratio. 
These regions which are close to, but not
coinciding with \HII~ regions have the same properties as similar
Galactic regions, and can be interpreted in the same way. They are probably
photodissociation interfaces. In these 
regions, the radiation field is probably strong enough to heat the very small 
interstellar grains to temperatures such that they contribute strongly
to the emission in the LW3 (12.0--18.0 $\mu$m) band.\\

\item Cesarsky et al. (\cite{Cesarsky98}) have observed a strange mid--IR
spectrum in four different regions of M\,31. It is characterized by a 
strong, broad emission band at 11.3 $\mu$m and faint bands at 6.2, 7.2 and 8.6 
$\mu$m compared to Galactic spectra. The present paper confirms this
property and shows that the regions that exhibit this particular type
of spectrum are only vaguely related with the regions emitting in the
LW2 and the LW3 filters, and are not related with the presence or absence of
UV radiation either.\\

\end{itemize}

The present results have profound consequences on our knowledge of the
mid--IR emission mechanisms in the interstellar medium. In particular,
we found that the mid-IR emission of the Galactic cirruses is not
dominated by UV excitation, but probably rather by visible photons, or
perhaps by another mechanism related to \HI~ that remains to be demonstrated.
It is only in regions submitted to a substantially higher UV radiation
field (with respect to the visible radiation field) that UV excitation
becomes important. The $\rho$ Ophiuchi cloud where the UV radiation 
density is about 10 times that in the Solar neighborhood (Abergel et al.
\cite{rhooph}) is such a region. 

The physical interpretation of our results is not easy and requires a 
detailed comparison of the properties of the mid--IR emission of the ISM in a 
wide variety of conditions of density, temperature and radiation field.
This will be the subject of a future paper, and we refrain here from any such
interpretation. However we wish to remark that the radiation field
in M\,31 is rather unique amongst nearby spiral galaxies by its deficiency 
in UV, due to its low star--formation rate even in the ``star--forming ring''
at 10 kpc from its center (M\,81 might be a similar case, but has been much 
less studied). The strange mid--IR properties of M\,31 (the
existence of spectra with a broad 11.3 $\mu$m band and little emission
in the 6.2, 7.7 and 8.6 $\mu$m bands) might be related to this lack of UV.
It will be of particular interest to look for 
mid--IR spectra or filter observations of other objects with little UV 
radiation, either in the archives of ISO or with future infrared space 
facilities. The central regions of M\,31 are still
poorer in UV than the region we have studied here. They are the suject
of a paper in preparation. We have seen that the bulge of another spiral 
galaxy, NGC 7331, has properties similar to those of M\,31 
(Smith \cite{Smith}). On the other hand, as discussed by Cesarsky
et al. (\cite{Cesarsky98}) the few available mid--IR spectra of elliptical
galaxies, which are other UV--poor objects containing ISM, look 
``normal'', i.e. similar to those of Galactic objects. This might be 
connected with the idea that their ISM comes from recent cannibalism of S 
or Irr galaxies in which the mid--IR emitters had ``normal'' properties, 
contrary to M\,31.\\ 

\noindent
{\bf Appendix: The interstellar radiation field in M\,31}

\noindent
From the visible surface photometry of M\,31 (Walterbos \& Kennicutt
\cite{Walterbos}) and our UV surface photometry at 200 nm, one can roughly 
estimate the spectral energy distribution of the ISRF in the ring of M\,31. 

The brightness in the V band in relatively unabsorbed regions of the 10 kpc 
ring is I$_{\rm V}$ $\simeq$ 22 mag. arcsec$^{-2}$ corresponding to
I$_{\lambda}$(550 nm) $\simeq$ 6 10$^{-18}$ erg cm$^{-2}$ s$^{-1}$
\AA$^{-1}$ arcsec$^{-2}$.  We estimate the UV brightness along the
same 21--cm line isophote of 1000 K km s$^{-1}$ on each side
of the 10 kpc ring. On the outer side (Fig. 4, yellow box 2) it is 1.2
10$^{-18}$ erg cm$^{-2}$ s$^{-1}$ \AA$^{-1}$ arcsec$^{-2}$ and on the
inner side (Fig. 4, yellow box 1) it is 0.3 10$^{-18}$ erg cm$^{-2}$
s$^{-1}$ \AA$^{-1}$ arcsec$^{-2}$. Hence I$_{\lambda}$(200
nm)/I$_{\lambda}$(550 nm) $\simeq$ 0.20 on the outer side and 0.05 on
the inner side. Note that extinction does not affect much the
UV/visible ratios at this column density. The column density of \HI~ is 1.835
10$^{21}$ H--atom cm$^{-2}$ corresponding to an total extinction
through the disk of 2.9 mag. at 220 nm and 1.1 mag. at 550 nm, assuming that 
the extinction law in M\,31 is the same as the Galactic one. The
contribution of molecular clouds to the average extinction is
small and is neglected here. If we assume
for simplicity that the UV sources have the same distribution
perpendicular to the plane as the dust responsible for extinction and
that the visible emission comes from a considerably thicker disk, we
calculate that the dereddened I$_{\lambda}$(220 nm)/I$_{\lambda}$(550
nm) ratios are only 1.55 times larger than the observed ones.

In the ``quiescent'' Field d of the ring (see Fig. 1) 
the V--band brightness is I$_{\rm V}$ $\simeq$ 21.75 mag. arcsec$^{-2}$
corresponding to I$_{\lambda}$(550 nm) = 7.5 10$^{-18}$ erg cm$^{-2}$ s$^{-1}$ 
\AA$^{-1}$ arcsec$^{-2}$. The average UV brightness (Fig. 4, yellow box
3) is 0.25 10$^{-18}$ erg cm$^{-2}$ s$^{-1}$ \AA$^{-1}$ arcsec$^{-2}$ .
Hence I$_{\lambda}$(200 nm)/I$_{\lambda}$(550 nm) $\simeq 0.03$. This
is similar to the ratio in the central 20\arcsec~ of M\,31 (Oke et al.
\cite{Oke}). The UV/visible ratio is probably more affected by interstellar 
extinction than in the previous regions. There is not only a
column density of 2.2 10$^{21}$ H--atom cm$^{-2}$ of \HI~ in Region 2 but
also a substantial amount of molecular hydrogen. The dereddening correction 
factor is uncertain but cannot be much larger than 3, thus the dereddened
I$_{\lambda}$(200 nm)/I$_{\lambda}$(550 nm) ratio is also small, probably
less than 0.1.

The I$_{\lambda}$(200 nm)/I$_{\lambda}$(550 nm) ratio 
for the local Galactic ISRF is $\simeq$ 0.66 from the 
calculations of Mathis et al. (\cite{Mathis}, Table A3). By direct integration 
of the B and I band starlight and interpolation to 550 nm (Tables 35 and 36 
in Leinert et al. \cite{Leinert}) and comparison with the ultraviolet ISRF 
of Gondhalekar et al. (\cite{Gondhalekar}, Fig. 6) we find 
I$_{\lambda}$(200 nm)/I$_{\lambda}$(550 nm) $\simeq$ 0.63, in excellent 
agreement with Mathis et al. (\cite{Mathis}). This is considerably larger 
than the above ratios for the M\,31 ring. This ring is thus deficient in 
UV with respect to the Solar neighbourhood, which itself is not an active
star--forming region. The ring of M 31 is even more deficient in UV
(with respect to the visible) than
the reflection nebula vdB 133, which is probably the most UV--deficient
Galactic object for which a mid--IR spectrum has been obtained. From the 
spectral classification 
and absolute magnitudes of the exciting stars given by Uchida and the 
color-color relations of Fig. 1 of Fanelli et al. (1987) we find a ratio
I$_{\lambda}$(200 nm)/I$_{\lambda}$(550 nm) $\simeq 0.4$, slightly
smaller than that in the Solar neighbourhood but definitively larger than
in the ring of M\,31.

The intensity of the ISRF in the visible is however of the same order in the 
ring of M\,31 and in the Solar neighbourhood. From Tables 35 and 36 in Leinert 
et al. (\cite{Leinert}) one can estimate that if seen from outside at 
the same inclination as M\,31 (77\degr), the Solar neighbourhood would have 
almost exactly the same V brightness as the M\,31 ring. The mean stellar 
brightness of the sky at galactic latitudes $b = \pm$13\degr~ is I$_{\rm V}$ 
$\simeq$ 22.2 mag. arcsec$^{-2}$. Thus the brightness of the Solar 
neighbourhood seen at an inclination of 77\degr~ would be:

I$_{\rm V}$(SN, 77\degr) $\simeq$ I$_{\rm V}$ + 
I$_{\rm V}$ (1 - exp(- A$_{\rm V}$)),

\noindent
where A$_{\rm V}$ is the extinction through the half--galactic plane at
latitude b = $\pm$13\degr. This extinction is about 1 magnitude from the
average \HI~ column density (Stark et al. \cite{Stark}), thus

I$_{\rm V}$(SN, 77\degr) $\simeq$ 21.8 mag. arcsec$^{-2}$,
 
\noindent
instead of 22 mag. arcsec$^{-2}$ for the ring of M\,31. As the UV/visible 
intensity ratio is 
smaller in M\,31, the absolute UV radiation density is correspondingly smaller
in M\,31 than in the Solar neighbourhood. This might have something to
do with
the difference between the ``normal'' mid--IR spectrum of the Galactic 
cirruses (Mattila et al. \cite{Mattila}; Onaka et al. \cite{Onaka}) and 
the strange mid--IR spectrum of M\,31.

\acknowledgements 
We thank Fran\c{c}ois Boulanger and Charly Ryter for illuminating discussions.
We also thank the referee for comments and questions which helped us
to improve the paper substantially.
\end{document}